# Preliminary studies of creation of gold nanoparticles on titanium surface towards biomedical applications

I. Csarnovics [a], P. Hajdu [b], S. Biri [b], Cs. Hegedüs [c], S. Kökényesi [a], R. Rácz [b], A . Csik [b]

[a] Institute of Physics, Department of Experimental Physics, University of Debrecen, H-4026 Debrecen, Bem tér 18/a, Hungary
[b] Institute for Nuclear Research (ATOMKI), Hungarian Academy of Sciences, H-4026 Debrecen, Bem tér 18/c, Hungary
[c] Department of Biomaterials and Prosthetic Dentistry, University of Debrecen, H-4012 Debrecen, Nagyerdei krt. 98, Hungary

## ABSTRACT

This paper is devoted to present the results of creation of gold nanoparticles on titanium surface. We focused on the problem how to create gold nanoparticles on the titanium surface with defined particle size and distribution, which could be interesting for several applications (e.g. providing well-defined substrates for biomedical research, etc.). To do that the sample is affected by the complex physical rout of gold nanoparticles formation: by gold ion implantation, thin Au layer deposition and thermal annealing. The effect of the technology, influence on the surface structure and its parameters were investigated by the X-ray diffraction, Scanning Electron and Atomic Force Microscopy, as well as by Secondary Neutral Mass Spectrometry methods.

Titanium is a widely used material for different applications in industry and medicine. The latter relates to implants with high stability and biocompatibility, where properties of the tissue-Ti/TiO$_2$ interaction are really important. Titanium itself is a biocompatible material; however its osseointegration can generally be influenced by modifying the surface structure, which is playing an important role in its successful clinical application [1,2]. Several studies show that the formation of a titanium oxide layer improves the biological compatibility of titanium and present many surface treatment methods to increasing the thickness of native oxide layer [3–5]. However, beside of this, the use of titanium implants with composite surfaces can exhibit antibacterial properties and influence cell growth processes [6,7]. These processes could be enhanced by adding gold nanoparticles (GNP), besides of that they also can easily establish special bonds to biomolecules [8]. A number of chemical and physical ways are known and used for GNP fabrication [8–13]. A preliminary study was done for surface modification of Ti by gold ions produced by electron cyclotron resonance (ECR) ion source (further called the ECR method) [14], but the created nanostructures were not analyzed deeply. In this work we present the results of structural changes investigated by different methods. One of the novelties of our works is improving the biocompatibility and osseointegration of Ti surface by gold ion beam produced by the ECR ion source (ECRIS) [15]. For the sample preparation we focused on two methods, used either separately or in combination:

(1) the nano-structurization of thin gold layer, deposited by magnetron sputtering method onto the surface of natural titanium sample kept under normal environmental conditions. We call it the physical vapor deposition (PVD) method.
(2) irradiation by Au ions provided by an ECR [15–17].

The heat treatment (annealing) of the samples also was applied in order to form GNP on the surface. Scanning Electron Microscope (SEM), Secondary Neutral Mass Spectrometry (SNMS) and Atomic Force Microscope (AFM) were used to establish the interconnection between the applied technology and the parameters of the resulted surface.

The Au ion beam was produced by the sputtering method [18], i.e. a gold pastille was bombarded by oxygen ions in the plasma. The extracted ion beam was straightly transported to the targets so all the beam components hit the samples. Before irradiation the extracted complex ion beam was analyzed by charge-to-mass ratio resulting in a beam spectrum. The peaks in the spectrum were identified and measured, and thus the ratios of the beam components could be exactly calculated. The composition of the beam was: Au − 2.4%, O − 78%, H − 10%, C − 8%, others 1.6%. The dose of the irradiation of gold ions was $1.5 \cdot 10^{16}$ ion/cm$^2$. Due to the different charges the gold ions were distributed not in a form of a thin layer, but stopped at different depth forming a density distribution in the surface layer of samples.

The 0.5 mm thick titanium plates (99.6 at%, grade 2, Spemet Co., Taipei, Taiwan) with dimension of 10 mm × 10 mm were mechanically polished to #2000 grid level, followed by 1 μm Al$_2$O$_3$ powder polishing to produce a mirror-like surface. All substrates were immersed in fresh 30% HNO$_3$ for 30 min at room temperature. In the next step all plates were sonicated in ethanol for 30 min, rinsing with distilled water for further 30 min and were dried in air.

Before ECR irradiation half part of the samples were covered by the stainless steel sheet. After irradiation this sheet was turned in 90 degree and the sample was covered with 15 nm thick gold layer by magnetron sputtering method. In this way we produced four different treated 5 × 5 mm parts on the same Ti sample. After preparation the samples were annealed at 550 °C for 6 h at atmospheric pressure to form gold nanoparticles from both Au components (irradiated and deposited). The following parts of the samples were investigated: 1 − pure Ti surface (Ti part), 2 − Ti surface irradiated with Au ions (ECR part), 3 − Ti surface irradiated with Au ions and covered with gold layer (ECR + PVD part), 4 − Ti surface covered with gold layer (PVD part).

Before and after annealing the samples were investigated with several different methods: SEM (Hitachi S-4300 CFE), AFM (Veeco diCaliber) and X-Ray diffraction (XRD). In order to have good statistical results all samples were investigated in 10 different places by SEM and AFM. The θ−2θ XRD measurements were carried out with a Siemens CuKα X-ray tube (λ = 1.54 Å) and a horizontal goniometer equipped with graphite monochromator. The parameters of the nanostructures were calculated using a standard image analyzing process on the SEM pictures made by National Instrument Vision Assistant. Energy dispersive x-ray spectroscopy (EDX) in the SEM system was used for checking the chemical components of the investigated samples. An INA-X type SNMS equipment (SPECS, Berlin) was used to measure the depth distribution of the elements to make a comparison with SRIM (*Stopping and Range of Ions in Matter*) simulation.

The implanted gold ions in Ti samples showed a Gaussian-like depth density distribution with a maximum around 10 nm in depth. The irradiation conditions were modeled by the well-known simulation code, SRIM [19]. Fig. 1 represents the comparison of the penetration calculation by SRIM to the data of the SNMS measurement, which was done on a sample before heat treatment. As it can be seen from Fig. 1(b) the gold ions penetrate into Ti up to about 13 nm, while oxygen strongly decreases in 3−5 nm depth showing the thickness of the oxide layer developed after ECR irradiation on Ti. The formation of this thin oxide layer (over the already existed ~25 nm thick natural titanium-oxide) probably also enhance the biological compatibility of the titanium implants [3,6]. In the SRIM simulations gold ions with realistic composition (obtained from the ECR beam spectra) were used. It was established that the model calculations are not perfect, but in good agreement with the experimental results; hence the SRIM can be used for modeling the penetration of the Au ions in the Ti samples during such a low-energy implantation. (For comparison Fig. 1(a) shows the theoretical distribution of gold ions in TiO, as well.) Fig. 1(b) shows that the implantation of Au ions was successful and the measured penetration depth corresponds with the SRIM calculation. It is also necessary to note that results of the SNMS method show that annealing (which is necessary for GNP creation) increases the thickness of the oxide layer on Ti surface up to 100÷120 nm.

Figs. 2 and 3 represent the SEM and AFM images of different parts of the Ti sample before and after heat treatment. AFM was used to measure the average roughness of the surface before and after the heat treatment. For the characterization of average roughness the root mean square (RMS) values for each samples was established from these measurements. The figures demonstrate the occurrence of nanostructures in case of the 4 areas due to annealing, whose parameters (average size and filling factor) depend on the surface treatment. The average roughness of different part of the samples, before heat treatment, was the following: Ti − 5 nm, Ti + ECR − 3 nm, Ti + ECR + PVD − 3 nm, Ti + PVD − 4 nm; while the roughness after heat treatment has changed: Ti − 11 nm, Ti + ECR − 9 nm, Ti + ECR + PVD − 17 nm, Ti + PVD − 25 nm. We can state that the heat treatment results in an increase of the roughness of the sample surface.

The structural characteristics of the pure Ti part of the sample were measured by XRD before and after annealing (Fig. 4). The spectrum of the investigated sample has changed definitely due to annealing. Appearance of TiO nanocrystals was observed after heat treatment. Characteristic peaks appeared at 35.1, 38.4, 40.2, 53.0 and 27.4, 35.9, 54.4 2θ degrees, which correspond to Ti and TiO$_2$ (rutile), respectively [20,21]. The size of the TiO$_2$ crystals has been calculated by the Scherrer-equation [22,23] and was determined in average size of 14 nm. The change of the roughness of pure Ti surface could be explained by the presence of TiO$_2$ nanostructures which is supported by XRD measurement. The XRD data is show − within the limits of sensitivity of this method − that before the annealing only metallic Ti exists, while $TiO_2$ nanocrystals appear after annealing. The peaks, which belong to these nanocrystals, are in connection with the rutile phase of the $TiO_2$.

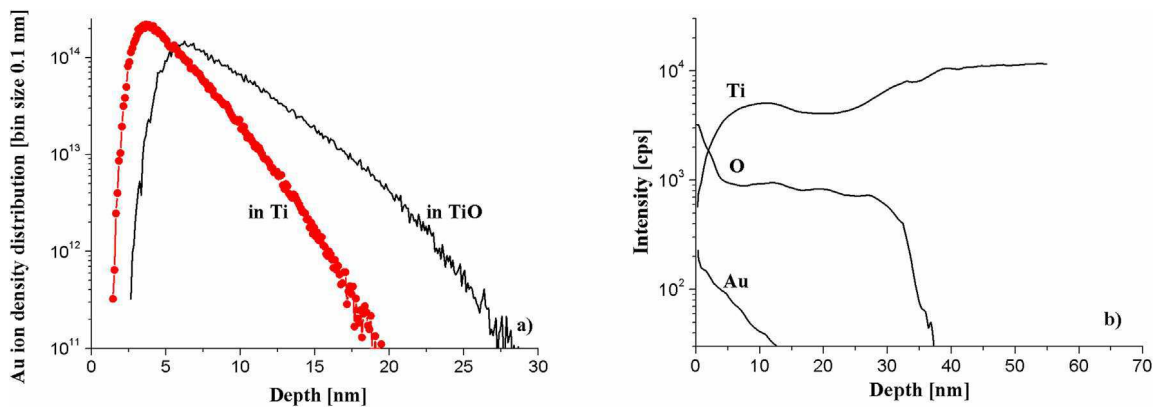

**Fig. 1.** a) Calculation with SRIM of the implantation of Au ions in the Ti and TiO samples. b) SNMS depth profile of the Ti sample, which was irradiated with Au ions before heat treatment.

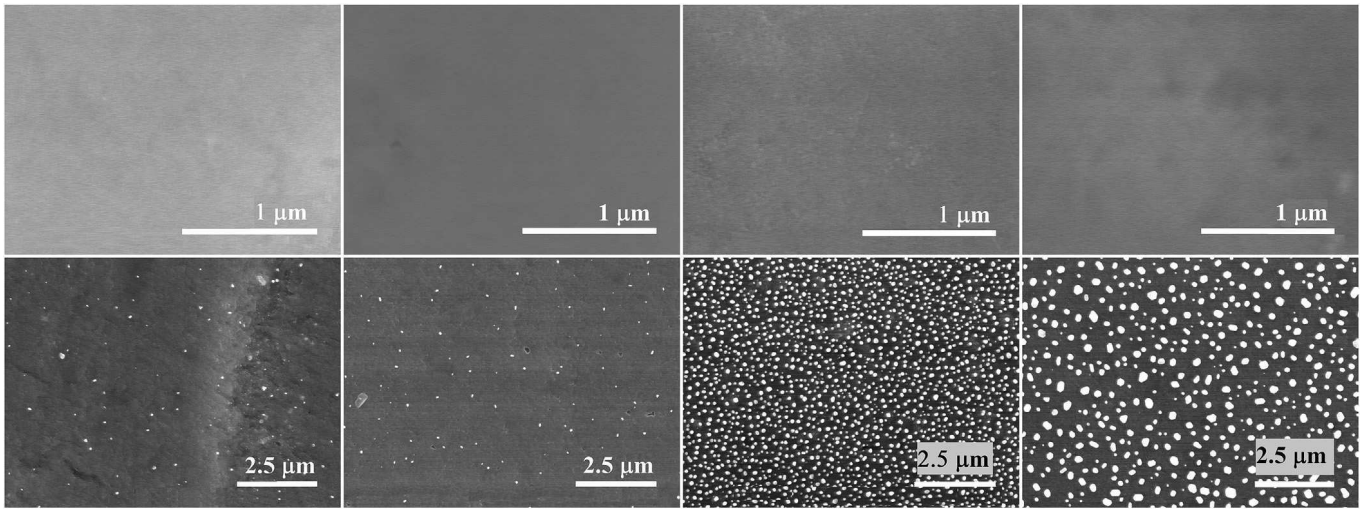

**Fig. 2.** SEM images of the surface of the Ti sample at different places. Upper pictures: before heat treatment, lower pictures: after heat treatment. From left to right: Ti part, ECR part, PVD part and ECR + PVD part.

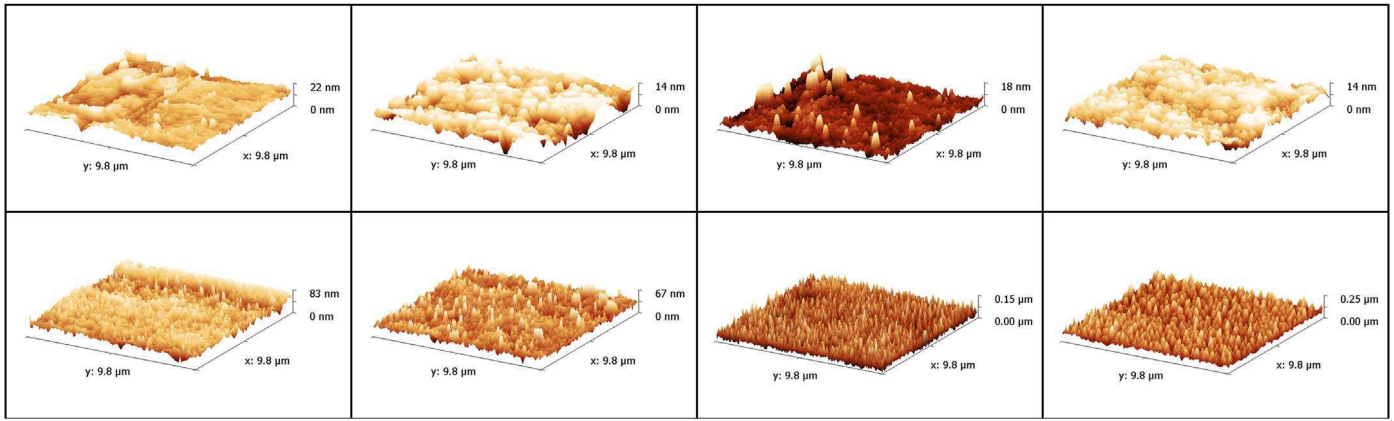

**Fig. 3.** AFM images of the surface of the Ti sample at different places. Upper pictures: before the heat treatment, lower pictures: after the heat treatment. From left to right: Ti part, ECR part, PVD part and ECR + PVD part.

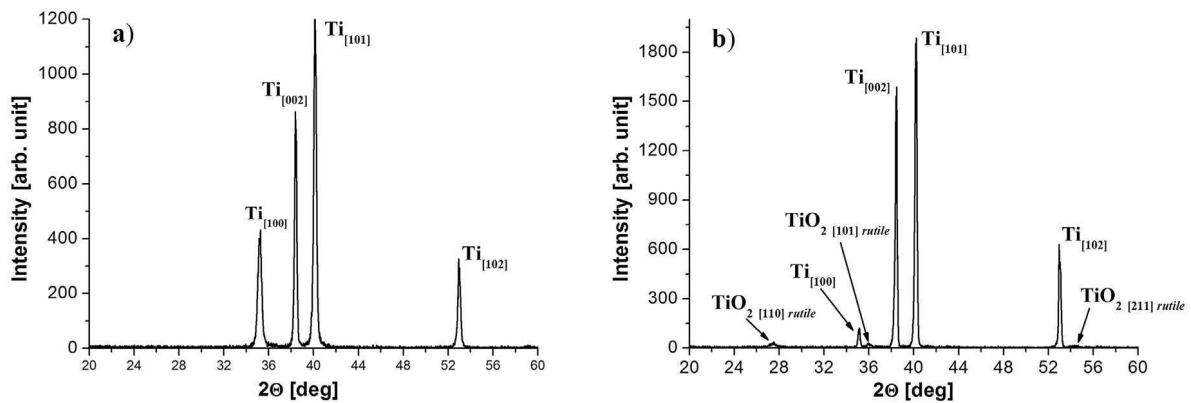

**Fig. 4.** X-ray diffraction patterns of the as-deposited (a) and annealed (b) pure Ti sample.

The average size (diameter) of the GNP on the sample after heat treatment were the following: Ti + ECR − 60 nm, Ti + ECR + PVD − 160 nm, Ti + PVD − 90 nm, while their filling factor: Ti + ECR − 0.3%, Ti + ECR + PVD − 11%, Ti + PVD − 11.2%. Analyzing the size distribution of the created particles we have determined that the size fit well with a Gaussian distribution (Fig. 5). The broader distribution in the case of ECR + PVD can be interpreted by the collective appearance of smaller and bigger GNPs. It seems that the smaller particles arise partially from the implanted Au, being partially embedded and increasing this way the mechanical stability of the nanocomposite. The appearance of bigger GNPs in the ECR + PVD part and the differences between the parameters of nanostructures can be explained by variation of diffusion rate and Ostwald-ripening on different parts of the sample. The surface of the Ti sample implanted with Au ions was less rough than without this treatment. On this part the process of Ostwald-ripening results a faster surface diffusion of gold. That's why the GNP creation was faster in the less rough part of the sample. Comparison of the ECR + PVD part with the PVD part showed us, that larger nanoparticles were formed, the filling factor and the number of the nanoparticles are smaller and the distance between the gold nanostructures is larger (Fig. 2). It is clearly observable that the relatively small amount of ECR-shot gold ions in the titanium remarkably changes the size distribution of GNPs created by PVD.

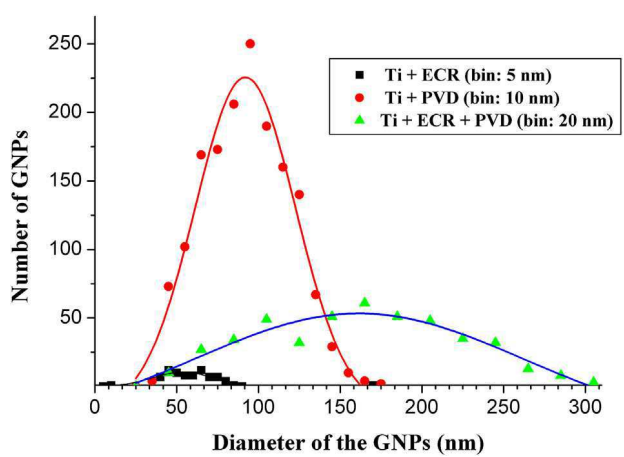

**Fig. 5.** The size distribution of GNPs (symbols). Lines are Gaussian curves.

Combined technology of titanium surface structurization, which includes formation of gold nanoparticles complemented with annealing was reported in this paper. We present a method for the formation of gold nanoparticles on titanium surface, created by the combination of different methods: ion implantation, deposition by magnetron sputtering method and thermal annealing. It was shown that the multiply charged Au ions were successfully implanted into Ti sample. The ion irradiation resulted in decreasing of the sample's surface roughness. Moreover, the depth profile of implanted gold proves that a small amount of gold nanostructures exist even under the surface of Ti, that can improve the stability of the resulting composite structure.

Gold nanoparticles of different dimensions were formed after gold layer deposition and annealing. In the same investigated area the amount of the detected gold particles was less in PVD than in ECR + PVD, which can be explained by the different size of the nanostructures (Fig. 5). Larger gold nanostructures were formed on that part which was previously implanted with Au ions (ECR + PVD) and they were complemented by smaller ones. In connection with this fact, the size of the created GNP on Ti can be easily controlled and investigated during the technology process.

Annealing of the sample results appearing of $TiO_2$ rutile nanocrystals which are biocompatible material and its combination with GNP can be used for further improvement of this type of implant materials. The biomedical effect of GNPs formed by the combined method shown in this paper is planned to be checked in the near future. In the case of bonding to biomolecules, a rough estimation can be given for the parameters (size and distance between nanoparticles) of the GNP system because these parameters are somehow related to the size of the biomolecules. Besides of that, it is planned to test the biomolecule immobilization on the GNP coated Ti surfaces.

### Acknowledgment

This work is supported by the Hungarian TÁMOP-4.2.2.A-11/1/KONV-2012-0036 project, which is co-financed by the European Union and the European Social Fund. The authors are grateful to Prof Dr. Shinn-Jyh Ding (Chung Shan Medical University, Institute of Oral Science, Taiwan) for the titanium samples.

### References

[1] F.Z. Cui, Z.S. Luo, Biomaterials modification by ion-beam processing, Surf. Coat. Technol. 112 (1999) 278–285.
[2] S. Nakano, M. Ishizuka, Q. Wang, H. Ogiso, Nanoparticle fabrication for micro total analysis system by using ion implantation and surface etching, Surf. Coat. Technol. 187 (2012) 167–171.
[3] K.L. Ou, Y.H. Shih, C.F. Huang, C.C. Chen, C.M. Liu, Preparation of bioactive amorphous-like titanium oxide layer on titanium by plasma oxidation treatment, Appl. Surf. Sci. 255 (2008) 2046–2051.
[4] I. Braceras, C. Vera, A. Ayerdi-Izquierdo, R. Munoz, J. Lorenzo, N. Alvarez, M. Ángel de Maeztu, Ion implantation induced nanotopography on titanium and bone cell adhesion, Appl. Surf. Sci. 310 (2014) 24–30.
[5] Y.-H. Lee, G. Bhattarai, N.-H. Lee, M.-H. Lee, T.-G. Kim, E.-C. Jhee, H-Y. Kim, H.-K. Yi, Modified titanium surface with gelatin nano gold composite increases osteoblast cell biocompatibility, Appl. Surf. Sci. 256 (2010) 5882–5887.
[6] L. Visai, L. De Nardo, C. Punta, L. Melone, A. Cigada, M. Imbriani, C.R. Arciola, Titanium oxide antibacterial surfaces in biomedical devices, Int. J. Artif. Organs 34 (2011) 929–946.
[7] T. Albrektsson, C. Johansson, Osteoinduction, osteoconduction and osseointegration, Eur. Spine. J. 10 (2001) 96–101.
[8] J.D. Padmos, P. Duchesne, M. Dunbar, P. Zhang, Gold nanoparticles on titanium and interaction with prototype protein, J. Biomed. Mater. Res. A 95 (2010) 146–155.
[9] J. Yang, J.Y. Lee, H.-P. Too, G.-M. Chow, L.M. Gan, Single stranded DNA stabilization and assembly of Au nanoparticles of different sizes, Chem. Phys. 323 (2006) 304–312.
[10] T. Arakawa, T. Kawahara, T. Akiyama, S. Yamada, Facile fabrication of gold nanoparticle-titanium oxide alternate assemblies by surface sol-gel process, J. Appl. Phys. 46 (2007) 48.
[11] J. Li, H. Zhou, S. Qian, Z. Liu, J. Feng, P. Jin, X. Liu, Plasmonic gold nanoparticles modified titania nanotubes for antibacterial application, Appl. Phys. Lett. 104 (2014) 261110.
[12] L. Tang, X. Guo, Y. Yang, Zh. Zha, Zh. Wang, Gold nanoparticles supported on titanium-dioxide: an efficient catalyst for highly selective synthesis of benzoxazoles and benzimidazoles, Chem. Commun. 50 (2014) 6145.
[13] M. Miljevic, B. Geiseler, T. Bergfeldi, P. Bockstaller, L. Fruk, Enhanced photocatalytic activity of Au/$TiO_2$ nanocomposite prepared using bifunctional bridging linker, Adv. Func. Mat. 24 (2014) 907–915.
[14] P. Hajdu, S. Biri, R. Racz, S. Kokenyesi, I. Csarnovics, A. Csik, K. Vad, Modification of titanium surface by gold ion beam, Acta Phys. Debr. XLVIII (2014) 1–12.
[15] S. Biri, I. Iván, Z. Juhász, B. Sulik, Cs. Hegedûs, A. Jenei, S. Kökényesi, J. Pálinkás, Application of the Atomki-ECRIS for materials research and prospects of the medical utilization, in: Proceedings of the 18th International Conf. on Electron cyclotron ion sources (ECRIS08), Chicago, IL, USA, 15-18 Sept 2008, pp. 39–45.
[16] S. Kökényesi, I. Iván, E. Takács, J. Pálinkás, S. Biri, A. Valek, Multipurpose 14.5 GHz ECR ion source: special features and application for surface modification, Nucl. Instr. Methods Phys. Res. B 223 (2005) 222–226.
[17] S. Biri, R. Rácz, J. Pálinkás, Status and special features of the Atomki ECR ion source, SCI Rev. Sci. Instrum. 83 (2012) 341.
[18] R. Rácz, S. Biri, I. Charnovych, S. Kökényesi, Gold and calcium ion beams for materials research by the ATOMKI ECR ion Source, Acta Phys. Deb. XLVI (2012) 133–141.
[19] J.F. Ziegler, SRIM-2003, Nucl. Instrum. Methods Phys. Res. B 219–220 (2004) 1027–1036.
[20] R.J. Wasilewski, Thermal expansion of Ti and some Ti-O-alloys, Trans. Met. Soc. AIME 221 (1961) 1231–1235.
[21] R. Restori, D. Schwarzenbach, J.R. Schneider, Charge density on rutile, $TiO_2$, Acta Crystallogr. Sec. B: Structural Science 43B (1987) 251–252.
[22] P. Scherrer, Bestimmung der gröse und der inneren struktur von kolloidteilchen mittels röntgenstrahlen, nachrichten von der gesellschaft der wissenschaften, Göttingen, Math.Phys. Kl. 2 (1918) 98.
[23] A. Patterson, The Scherrer formula for X-Ray particle size determination, Phys. Rev. 56 (1939) 978–982.